\documentclass[12pt]{article}
\usepackage{graphicx}
\usepackage{lscape}
\usepackage{citesort}
\usepackage{amssymb}
\usepackage{appendix}
\usepackage{multirow}
\usepackage[table]{xcolor}
\usepackage{colortbl}
\definecolor{lightgray}{gray}{0.9}

\usepackage{mathrsfs}
\begin{document}
\def\qq{\langle \bar q q \rangle}
\def\uu{\langle \bar u u \rangle}
\def\dd{\langle \bar d d \rangle}
\def\sp{\langle \bar s s \rangle}
\def\GG{\langle g_s^2 G^2 \rangle}
\def\Tr{\mbox{Tr}}
\def\figt#1#2#3{
        \begin{figure}
        $\left. \right.$
        \vspace*{-2cm}
        \begin{center}
        \includegraphics[width=10cm]{#1}
        \end{center}
        \vspace*{-0.2cm}
        \caption{#3}
        \label{#2}
        \end{figure}
    }

\def\figb#1#2#3{
        \begin{figure}
        $\left. \right.$
        \vspace*{-1cm}
        \begin{center}
        \includegraphics[width=10cm]{#1}
        \end{center}
        \vspace*{-0.2cm}
        \caption{#3}
        \label{#2}
        \end{figure}
                }

\def\ds{\displaystyle}
\def\beq{\begin{equation}}
\def\eeq{\end{equation}}
\def\bea{\begin{eqnarray}}
\def\eea{\end{eqnarray}}
\def\beeq{\begin{eqnarray}}
\def\eeeq{\end{eqnarray}}
\def\ve{\vert}
\def\vel{\left|}
\def\ver{\right|}
\def\nnb{\nonumber}
\def\ga{\left(}
\def\dr{\right)}
\def\aga{\left\{}
\def\adr{\right\}}
\def\lla{\left<}
\def\rra{\right>}
\def\rar{\rightarrow}
\def\lrar{\leftrightarrow}
\def\nnb{\nonumber}
\def\la{\langle}
\def\ra{\rangle}
\def\ba{\begin{array}}
\def\ea{\end{array}}
\def\tr{\mbox{Tr}}
\def\ssp{{\Sigma^{*+}}}
\def\sso{{\Sigma^{*0}}}
\def\ssm{{\Sigma^{*-}}}
\def\xis0{{\Xi^{*0}}}
\def\xism{{\Xi^{*-}}}
\def\qs{\la \bar s s \ra}
\def\qu{\la \bar u u \ra}
\def\qd{\la \bar d d \ra}
\def\qq{\la \bar q q \ra}
\def\gGgG{\la g^2 G^2 \ra}
\def\q{\gamma_5 \not\!q}
\def\x{\gamma_5 \not\!x}
\def\g5{\gamma_5}
\def\sb{S_Q^{cf}}
\def\sd{S_d^{be}}
\def\su{S_u^{ad}}
\def\sbp{{S}_Q^{'cf}}
\def\sdp{{S}_d^{'be}}
\def\sup{{S}_u^{'ad}}
\def\ssp{{S}_s^{'??}}

\def\sig{\sigma_{\mu \nu} \gamma_5 p^\mu q^\nu}
\def\fo{f_0(\frac{s_0}{M^2})}
\def\ffi{f_1(\frac{s_0}{M^2})}
\def\fii{f_2(\frac{s_0}{M^2})}
\def\O{{\cal O}}
\def\sl{{\Sigma^0 \Lambda}}
\def\es{\!\!\! &=& \!\!\!}
\def\ap{\!\!\! &\approx& \!\!\!}
\def\md{\!\!\!\! &\mid& \!\!\!\!}
\def\ar{&+& \!\!\!}
\def\ek{&-& \!\!\!}
\def\kek{\!\!\!&-& \!\!\!}
\def\cp{&\times& \!\!\!}
\def\se{\!\!\! &\simeq& \!\!\!}
\def\eqv{&\equiv& \!\!\!}
\def\kpm{&\pm& \!\!\!}
\def\kmp{&\mp& \!\!\!}
\def\mcdot{\!\cdot\!}
\def\erar{&\rightarrow&}
\def\olra{\stackrel{\leftrightarrow}}
\def\ola{\stackrel{\leftarrow}}
\def\ora{\stackrel{\rightarrow}}
% .........................................................

\def\simlt{\stackrel{<}{{}_\sim}}
\def\simgt{\stackrel{>}{{}_\sim}}

% .........................................................

\title{
         {\Large
                 {\bf
                   Semileptonic $B \rightarrow \bar{D} $ transition in
nuclear medium
                 }
         }
      }

\author{\vspace{1cm}\\
{\small  K. Azizi$^1$ \thanks {e-mail: kazizi@dogus.edu.tr}\,\,, N. Er$^2$\thanks {e-mail: nuray@ibu.edu.tr}, H.
Sundu$^3$\thanks {e-mail: hayriye.sundu@kocaeli.edu.tr}} \\
{\small $^1$ Department of Physics, Do\u gu\c s University,
Ac{\i}badem-Kad{\i}k\"oy, 34722 \.{I}stanbul, Turkey}\\
{\small $^2$ Department of Physics, Abant \.{I}zzet Baysal University,
G\"olk\"oy Kamp\"us\"u, 14980 Bolu, Turkey}\\
{\small $^3$ Department of Physics, Kocaeli University,
 41380 \.{I}zmit, Turkey}\\
}
\date{}

\begin{titlepage}
\maketitle
\thispagestyle{empty}
\begin{abstract}
We study the semileptonic tree-level  $B \rightarrow \bar{D}$ transition   in the framework of  QCD sum rules in nuclear medium. In particular, we calculate the in-medium form factors entering the transition matrix elements
defining this decay channel. It is found that the interactions of the participating particles with the medium lead to  a considerable suppression in the branching ratio compared to the vacuum.

\end{abstract}
PACS number(s):13.20.He, 21.65.Jk, 11.55.Hx
\end{titlepage}

% The aim of this article
% Standard model and new physics
% Topcolor-assisted technicolor model

                                 %%%%%%%%%%%%%%%%%%%%%%%%%%%%%%%%%%%%%%%
       %%%%%%%%%%%%%%%%%%%%%%%%%%%%%%%%%%%%%%%             %%%%%%%%%%%%%%%%%%%%%%%%%%%%%%%%%%%%%%%%%%
                                 %%%%%%%%%%%%%%%%%%%%%%%%%%%%%%%%%%%%%%%
\section{Introduction}

It is well-known that the semileptonic $B$ meson decay channels are excellent frameworks to  calculate the
standard model  parameters, confirm its validity,  understand the origin of  CP
violation and search for new physics effects. There are many theoretical and experimental studies devoted to the  semileptonic tree-level  $B \rightarrow D$ transition  in vacuum for many years. After 
the BABAR \cite{BABAR} measurement  of the ratio of branching fractions in $\tau$ and $\ell=\mu, e$ channels, i.e.
%Among the
%$B$-meson decay channels, the $B\rightarrow D \ell
%\overline{\nu}_{\ell}$ tansition is the most important since it is
%used to determine the value of the Cabibbo-Kobayashi-Maskawa (CKM)
%matrix $\mid V_{cb} \mid$ and the $D$-meson distribution
%amplitude.
$\ensuremath{{\cal R}(D)} ={\cal B}(\overline{B}\rightarrow D
\tau^{-}\overline{\nu}_{\tau})/{\cal B}(\overline{B}\rightarrow D
\ell^{-}\overline{\nu}_{\ell})=0.440\pm 0.058\pm 0.042$ which deviates with the standard model expectations
with a $3.4~\sigma$, this channel together with  a similar anomaly in $B \rightarrow D^*$ transition have raised interests to study these channels 
via different models (for some of them see \cite{colangelo,fajfer,fajfer1,Crivellin,Datta,Becirevic,damir1,Celis,Choudhury,Tanaka,Bhol} and references therein).

%anomaly,
%which exceed the SM expectations by $2.0\sigma$ \cite{BABAR}, the
%semileptonic $B$ to $D$ meson decays have been in the focus of much
%attention both theoretically and experimentally.

%In the literature, there are many experimental and theoretical
%works devoted to the study of the $B$ meson decay channels. 
%BABAR
%Collaboration report a measurement of the exclusive charmless
%semileptonic decay $B^+\rightarrow \omega \ell^+ \nu$
%\cite{BABAR1}. LHCb Collaboration studies the decay of the
%$B^0\rightarrow K^{*0}\mu^+\mu^-$ \cite{LHCb,LHCb1}. In the QCD
%sum rules method, $B_s(B^{\pm})(B^0)\rightarrow
%D_s[1968](D^0)(D^{\pm})\ell\nu$ decays are investigated
 %at vacuum \cite{Azizi}. In order to determine $\mid V_{cb} \mid$ and the
%properties of the $D$-meson distribution amplitude, $B\rightarrow
%D \ell\overline{\nu_{\ell}}$ is studied in the framework of QCD
%light-cone sum rule (LCSR) \cite{Fu}. The $B\right arrow
%D_{2}^{*}(2460)\ell\overline{\nu}$ is investigated within the
%framework of the three-point QCD sum rules \cite{Azizi1}. In the
%\cite{Segovia,Segovia1,Fu1,Ivanov}, the semileptonic, nonleptonic
%and rare decay modes of $B$ meson are analyzed. In \cite{Katirci},
%$B\rightarrow K_{2}^{*}(1430)\ell^+\ell^-$ is investigated in
%universal extra dimension model in order to search new physics
%beyond the SM. Also, in \cite{Damir}, new physics is search via
%$\overline{B}\rightarrow D\tau\overline{\nu}_{\tau}$ and
%$\overline{B}\rightarrow D\mu\overline{\nu}_{\mu}$.

The in-medium studies on the spectroscopic properties of the $B$ and $D$ mesons \cite{Hayashigaki,Azizi2,Hilger,Hilger2,Wang2011} show that the masses and decay constants
of these mesons receive modifications from the interactions of these particles with the nuclear medium.
It is expected that the form factors governing the semileptonic   $B $ to $ D$ transition are also affected by these interactions. In this accordance we calculate the in-medium 
transition form factors entering the  low energy matrix elements defining the  semileptonic tree-level  $B\rightarrow \bar{D}$ transition in the framework of the QCD sum rules. 
 This is the first attempt to calculate the hadronic transition 
form factors in the nuclear medium. Using  the transition form factors, we also calculate the decay width and branching ratio of this transition in nuclear medium. 
Study the in-medium properties of hadrons and their decays can help us in better  understanding the perturbative and non-perturbative natures of QCD. This   can also
play crucial role in analyzing the results of heavy ion collision experiments held at different places. There have been a lot of experiments such as CEBAF and RHIC  focused on the study 
of the hadronic  properties in nuclear medium. The FAIR and CBM Collaborations intend
to study the in-medium properties of different hadrons. The PANDA Collaboration also plans to focus on the study of the charmed hadrons \cite{bir,iki,uc,dort}.
We hope it will be possible to experimentally study the in-medium properties of the decay channels like $B\rightarrow \bar{D}$ transition  in near future.

%in-medium and calculate some related physical quantities within
%the framework of the three point QCD sum rules. Although there are
%some theoretical works devoted to study of the in-medium
%properties of $B$ meson \cite{Hayashigaki,Azizi2}, its decay
%channels properties in nuclear matter are not studied in
%literature. In this regard, this work can be useful for the
%literature. Also, our results can be used in analysis of the
%obtained by heavy ion collisions experiments.

The article is organized as follows. Next section 
includes the details of calculations of the transition form factors for
the semileptonic tree-level $B\rightarrow \bar{D} $ in nuclear medium via  QCD sum
rules. In section 3, we present  our numerical analysis of the form
factors and estimate  the  branching ratio of
the decay channel under consideration.

\section{In-medium transition form factors}

The $B^{+}\rightarrow \overline{D}^{0} \ell'^{+}
\nu_{\ell'}$  decay, where $\ell'^{+}$ can be either $\ell^{+}$=(e, $\mu$) or $\tau$,  proceeds via $ \overline{b}\rightarrow  \overline{c}\ell'^{+}
\nu_{\ell'}$    transition at quark level whose effective Hamiltonian can be written as
\begin{equation}
\label{matrixelement}
H_{eff}=\frac{G_F}{\sqrt{2}}V_{cb}\nu_{\ell'}\gamma_{\mu}(1-\gamma_5)\ell'^{+} c\gamma_{\mu}(1-\gamma_5)\overline{b},
\end{equation}
where $G_F$ is the Fermi coupling constant and $V_{cb}$ is an element of the Cabibbo-Kobayashi-Maskawa (CKM) matrix.
The amplitude of this transition is given as
\begin{equation}
\label{amplitude}
M=\frac{G_F}{\sqrt{2}}V_{cb}\nu_{\ell'}\gamma_{\mu}(1-\gamma_5)\ell'^{+} \langle  \overline{D}^{0}(p')| c\gamma_{\mu}(1-\gamma_5)\overline{b} |B^+(p) \rangle,
\end{equation}
where, to proceed, we shall define  the matrix element $\langle  \overline{D}^{0}(p')| c\gamma_{\mu}(1-\gamma_5)\overline{b} |B^+(p) \rangle$ in terms of
transition form factors. The transition current consists of axial vector  and vector parts. 
The first one has no contribution due to the parity and Lorentz considerations, but the second one can be parametrized in terms of two transition form factors $f_1(q^2)$ and $f_2(q^2)$ in the following way:
\begin{equation}
\label{Pmuqmu}
\langle   \overline{D}^{0}(p')| c\gamma_{\mu}\overline{b} |B^+(p) \rangle=f_1(q^2)P_{\mu}+f_2(q^2)q_{\mu}
\end{equation}
where $P=p+p'$ and  $q=p-p'$.

 In order to calculate the form factors, the  following in-medium three-point correlation function is considered:
\begin{equation}\label{corre}
\Pi_\mu(q^2)  =  i^2 \int d^4 x d^4 y e^{-ip\cdot x} e^{ip' \cdot y} \langle \psi_0 |{\cal T}[J_{\overline{D}^{0}}(y)J_{\mu}^{tr}(0)J_{B^+}^\dag(x)] | \psi_0 \rangle
\end{equation}
where $|\psi_0\rangle$ is the nuclear matter ground state, ${\cal T}$ is the time-ordering operator, $J_{\mu}^{tr}(0)$ is the transition current; and $J_{\overline{D}^{0}} (y) = \bar{c} (y)i \gamma_5 u(y)$ and  $J_{B^+} (x) = \bar{b} (x) i\gamma_5 u(x)$ are interpolating currents of  the $\overline{D}^{0}$  and $B^+$ mesons,
 respectively.

We shall calculate this correlator in two different ways: in terms of the in-medium hadronic parameters called
 the hadronic side and in terms of the in-medium QCD degrees of freedom defining in terms of nuclear matter density using the operator product expansion (OPE)   called the OPE  side. 
By equating these two representations to each other, the 
 in-medium form factors  are obtained. To suppress the contributions of the higher states and continuum, we apply Borel transformation and continuum subtraction to both sides of the sum rules obtained and use the quark-hadron duality assumption.

\subsection{Hadronic side}

On the  hadronic side,   the correlation function in Eq.(\ref{corre}) is calculated via implementing  two complete sets of intermediate states with the same quantum numbers as the currents $J_{D}$ and $J_{B}$.
 After performing the four-integrals we get
\begin{equation}
\label{hadronic}
\Pi^{HAD}_{\mu} =\frac{\langle \psi_0|J_{\overline{D}^{0}}(0)|\overline{D}^{0}(p')\rangle \langle \overline{D}^{0}(p')|J_{\mu}^{tr}(0)|B^+(p)\rangle_{\psi_0} \langle B^+(p)|J_{B^+}^\dag(0)|\psi_0\rangle}{(p'^2-m^2_{D^*})(p^2-m^2_{B^{*}})}+...,
\end{equation}
where the dots denote the contributions coming from the  higher states and continuum. We have previously defined the transition matrix elements in terms of form factors. The 
remaining matrix elements in the above equation are defined as
\begin{eqnarray}
\label{matrixelem}
\langle \psi_0|J_{\overline{D}^{0}}(0)|\overline{D}^{0}(p')\rangle&=&i\frac{f^*_{D}m^{*2}_{D}}{m_c+m_u}, \nonumber \\
 \langle B^+(p)|J^\dag_{B^+}(0)|\psi_0\rangle&=&-i\frac{f^*_{B}m^{*2}_{B}}{m_b+m_u},
\end{eqnarray}
where  $m^*_{D}$, $m^*_{B}$, $f^*_{D}$ and $f^*_{B}$ are the masses and the leptonic decay constants of $D$ and $B$ mesons in nuclear medium. These quantities in nuclear matter are calculated in
 \cite{Azizi2}.  Using Eqs.(\ref{Pmuqmu}) and (\ref{matrixelem}), one can write  Eq.(\ref{hadronic}) in terms of two different structure as 
\begin{equation}
\label{structure}
\Pi^{HAD}_{\mu}(q^2)=\Pi_{1}(q^2)P_{\mu}+\Pi_{2}(q^2)q_{\mu}
\end{equation}
where
\begin{equation}
\label{pi1}
\Pi_{1}(q^2)=-\frac{1}{(p'^2-m^{*2}_{D})(p^2-m^{*2}_{B})}\frac{f^*_{D}m^{*2}_{D}}{m_c+m_u}\frac{f^*_{B}m^{*2}_{B}}{m_b+m_u}f_1(q^2)+...,
\end{equation}
and
\begin{equation}
\label{pi2}
\Pi_{2}(q^2)=-\frac{1}{(p'^2-m^{*2}_{D})(p^2-m^{*2}_{B})}\frac{f^*_{D}m^{*2}_{D}}{m_c+m_u}\frac{f^*_{B}m^{*2}_{B}}{m_b+m_u}f_2(q^2)+....
\end{equation}

\subsection{OPE side}

The OPE side of the correlation function is calculated via inserting  the explicit forms of the interpolating currents into Eq. (\ref{corre}).  After contracting out all quark pairs using the Wick's 
theorem we get
\begin{equation}
\label{corOPE}
\Pi^{OPE}_{\mu}(q^2)=(-1)^3 i^4 \int d^4 x d^4 y e^{-ip\cdot x} e^{ip' \cdot y} \langle \psi_0 |Tr[S_u(y-x)\gamma_5 S_b(x)\gamma_\mu (1-\gamma_5) S_c(-y)\gamma_5  | \psi_0 \rangle,
\end{equation}
where $S_q$ with $q=u$ and $S_Q$ with $Q=b$ or $ c$ are the light and heavy quark propagators. In
coordinate-space the light quark
propagator at  nuclear medium and in the fixed-point gauge is given by
\cite{cohen95,reinders85}
\begin{eqnarray}\label{propagatorlight}
 S_{q}^{ab}(x)&\equiv& \langle
 \psi_0|{\cal T}[q^a
(x)\bar{q}^b(0)]|\psi_0\rangle_{\rho_N}\nonumber \\
&=&
\frac{i}{2\pi^2}\delta^{ab}\frac{1}{(x^2)^2}\not\!x
-\frac{m_q }{ 4\pi^2} \delta^ { ab } \frac { 1}{x^2} +
\chi^a_q(x)\bar{\chi}^b_q(0)-\frac{ig_s}{32\pi^2}F_{\mu\nu}^A(0)t^{ab,A
}\frac{1}{x^2}[\not\!x\sigma^{\mu\nu}+\sigma^{\mu\nu}\not\!x]\nonumber \\&+&...,
\end{eqnarray}
where $\rho_N$ is the nuclear matter density. The first and second terms on the right-hand side denote the
expansion of the free quark propagator
to first order in the light quark mass (perturbative part); and the third and forth
terms represent the contributions due to the background quark and gluon fields
(non-perturbative part). The heavy quark propagator is also taken as
\begin{eqnarray}\label{propagatorheavy}
 S_{Q}^{ab}(x)&\equiv& \langle\psi_0|{\cal T}[Q^a
(x)\bar{Q}^b(0)]|\psi_0\rangle_{\rho_N} \nonumber \\
&=&
\frac{i}{(2\pi)^4}\int d^4 k e^{(-ik\cdot x)}\Bigg\{
\frac{\delta^{ab}}{\not\!k-m_Q}-\frac{g_s
G^n_{\alpha\beta}t_{ab}^n}{4}\frac{\sigma^{\alpha\beta}
(\not\!k+m_Q)+(\not\!k+m_Q)\sigma^{\alpha\beta}}{(k^2-m_Q^2)^2}\nonumber
\\&&+\frac{ \delta_{ab} \langle g_s^2 GG\rangle }{12}\frac{m_Qk^2+m_Q^2
\not\!k}{(k^2-m_Q^2)^4}+... \Bigg\} ,
\end{eqnarray}
where  $t^n=\frac{\lambda^n}{2}$ with  $\lambda^n$ 
being the Gell-Mann
matrices. 

 The next step is to use   Eqs.  (\ref{propagatorlight}) and
 (\ref{propagatorheavy}) in Eq. (\ref{corOPE}) and define the following operators   
\begin{eqnarray}\label{}
&&\chi_{a\alpha}^{q}(x)\bar{\chi}_{b\beta}^{q}(0)=\langle
q_{a\alpha}(x)\bar{q}_{ b\beta}(0)\rangle_{\rho_N}, ~~~~~~~
F_{\kappa\lambda}^{A}F_{\mu\nu}^{B}=\langle
G_{\kappa\lambda}^{A}G_{\mu\nu}^{B}\rangle_{\rho_N}, \nonumber \\
&&\chi_{a\alpha}^{q}\bar{\chi}_{b\beta}^{q}F_{\mu\nu}^{A}=\langle q_{a\alpha}\bar{q}_{ b\beta}G_{\mu\nu}^{A}\rangle_{\rho_N},
~~~~~~~~\chi_{a\alpha}^{q}\bar{\chi}_{b\beta}^{q}\chi_{c\gamma}^{q}\bar
{\chi}_{d\delta}^{q}=\langle
q_{a\alpha}\bar{q}_{b\beta}
q_{c\gamma}\bar{q}_{d\delta}\rangle_{\rho_N}.
\end{eqnarray}

 The matrix element $\langle
q_{a\alpha}(x)\bar{q}_{b\beta}(0)\rangle_{\rho_N}$ is expanded as \cite{cohen95}
\begin{eqnarray} \label{ }
\angle
q_{a\alpha}(x)\bar{q}_{b\beta}(0)\rangle_{\rho_N}&=&-\frac{\delta_{ab}}{12}\Bigg
[\Bigg(\langle\bar{q}q\rangle_{\rho_N}+x^{\mu}\langle\bar{q}D_{\mu}q\rangle_{
\rho_N}
+\frac{1}{2}x^{\mu}x^{\nu}\langle\bar{q}D_{\mu}D_{\nu}q\rangle_{\rho_N}
+...\Bigg)\delta_{\alpha\beta}\nonumber \\
&&+\Bigg(\langle\bar{q}\gamma_{\lambda}q\rangle_{\rho_N}+x^{\mu}\langle\bar{q}
\gamma_{\lambda}D_{\mu} q\rangle_{\rho_N}
+\frac{1}{2}x^{\mu}x^{\nu}\langle\bar{q}\gamma_{\lambda}D_{\mu}D_{\nu}
q\rangle_{\rho_N}
+...\Bigg)\gamma^{\lambda}_{\alpha\beta} \Bigg].\nonumber \\
\end{eqnarray}
The quark-gluon mixed condensate in nuclear matter is written as
\begin{eqnarray} \label{ }
\langle
g_{s}q_{a\alpha}\bar{q}_{b\beta}G_{\mu\nu}^{A}\rangle_{\rho_N}&=&-\frac{t_{
ab}^{A
}}{96}\Bigg\{\langle g_{s}\bar{q}\sigma\cdot
Gq\rangle_{\rho_N}\Bigg[\sigma_{\mu\nu}+i(u_{\mu}\gamma_{\nu}-u_{\nu}\gamma_{\mu
})
\!\not\! {u}\Bigg]_{\alpha\beta} \nonumber \\
&&+\langle g_{s}\bar{q}\!\not\! {u}\sigma\cdot
Gq\rangle_{\rho_N}\Bigg[\sigma_{\mu\nu}\!\not\!
{u}+i(u_{\mu}\gamma_{\nu}-u_{\nu}\gamma_{\mu}
)\Bigg]_{\alpha\beta} \nonumber \\
&&-4\Bigg(\langle\bar{q}u\cdot D u\cdot D q\rangle_{\rho_N}+im_{q}\langle\bar{q}
\!\not\! {u}u\cdot D q\rangle_{\rho_N}\Bigg) \nonumber \\
&&\times\Bigg[\sigma_{\mu\nu}+2i(u_{\mu}\gamma_{\nu}-u_{\nu}\gamma_{\mu}
)\!\not\! {u}\Bigg]_{\alpha\beta}\Bigg\},
\end{eqnarray}
where  $D_\mu=\frac{1}{2}(\gamma_\mu \!\not\!{D}+\!\not\!{D}\gamma_\mu)$ and $u_{\mu}$ is the four velocity vector of the nuclear medium.
The matrix element of the four-dimension gluon condensate can also be  written as
\begin{equation}
 \langle
G_{\kappa\lambda}^{A}G_{\mu\nu}^{B}\rangle_{\rho_N}=\frac{\delta^{AB}}{96}
\Bigg[
\langle
G^{2}\rangle_{\rho_N}(g_{\kappa\mu}g_{\lambda\nu}-g_{\kappa\nu}g_{\lambda\mu}
)+O(\langle
\textbf{E}^{2}+\textbf{B}^{2}\rangle_{\rho_N})\Bigg],
\end{equation}
where we neglect the last term in this equation due to its small contribution. We also 
ignore from the four-quark condensate contributions in our calculations.
The above equations  contain various condensates, which are  are defined as \cite{cohen95,XJ1}
\begin{eqnarray} \label{ }
\langle\bar{q}\gamma_{\mu}q\rangle_{\rho_N}&=&\langle\bar{q}\!\not\!{u}q\rangle_
{\rho_N}
u_{\mu} , \\
\langle\bar{q}D_{\mu}q\rangle_{\rho_N}&=&\langle\bar{q}u\cdot D
q\rangle_{\rho_N}
u_{\mu}=-im_{q}\langle\bar{q}\!\not\!{u}q\rangle_{\rho_N}
u_{\mu}  ,\\
\langle\bar{q}\gamma_{\mu}D_{\nu}q\rangle_{\rho_N}&=&\frac{4}{3}\langle\bar{q}
\!\not\! {u}u\cdot D q\rangle_{\rho_N}(u_{\mu}u_{\nu}-\frac{1}{4}g_{\mu\nu})
+\frac{i}{3}m_{q} \langle\bar{q}q\rangle_{\rho_N}(u_{\mu}u_{\nu}-g_{\mu\nu}),
\\
\langle\bar{q}D_{\mu}D_{\nu}q\rangle_{\rho_N}&=&\frac{4}{3}\langle\bar{q}
u\cdot D u\cdot D q\rangle_{\rho_N}(u_{\mu}u_{\nu}-\frac{1}{4}g_{\mu\nu})
-\frac{1}{6} \langle
g_{s}\bar{q}\sigma\cdot Gq\rangle_{\rho_N}(u_{\mu}u_{\nu}-g_{\mu\nu}) ,\\
\langle\bar{q}\gamma_{\lambda}D_{\mu}D_{\nu}q\rangle_{\rho_N}&=&2\langle\bar{q}
\!\not\! {u}u\cdot D u\cdot D
q\rangle_{\rho_N}\Bigg[u_{\lambda}u_{\mu}u_{\nu} -\frac{1}{6}
(u_{\lambda}g_{\mu\nu}+u_{\mu}g_{\lambda\nu}+u_{\nu}g_{\lambda\mu})\Bigg]\nonumber\\
&&-\frac{1}{6} \langle
g_{s}\bar{q}\!\not\! {u}\sigma\cdot
Gq\rangle_{\rho_N}(u_{\lambda}u_{\mu}u_{\nu}-u_{\lambda}g_{\mu\nu}),
\end{eqnarray}
where the  equations of motion have been used and $\textit{O}(m^2_q)$ terms have also  been ignored due to their very small contributions.

The correlation function on OPE side can be written in terms of  the perturbative and non-perturbative parts  as
\begin{eqnarray}
\label{piope}
\Pi^{OPE}_{\mu}(q^2)&=&\Big[\Pi^{pert}_{1}(q^2)+\Pi^{non-pert}_{1}(q^2)\Big]P_{\mu} \nonumber \\
&+&\Big[\Pi^{pert}_{2}(q^2)+\Pi^{non-pert}_{2}(q^2)\Big]q_{\mu},
\end{eqnarray}
where
\begin{equation}
\label{Pipert12}
\Pi^{pert}_{1[2]}(q^2)=\int ds \int ds' \frac{\rho_{1[2]}( s,s',q^2)}{(s-p^2)(s'-p^{'2})} + \textrm{subtraction terms}.
\end{equation}
After some lengthy but straightforward calculations, the spectral densities $\rho_{1[2]}(s,s',q^2)$ are obtained as
\begin{eqnarray}
\rho_1( s,s',q^2) & = & \int_0^1 dz \int_0^{1-z} dw \frac{-3(5w+5z-4)}{8\pi^2}\theta[L(s,s',q^2)], \\
\rho_2( s,s',q^2) & = & \int_0^1 dz \int_0^{1-z} dw \frac{15(w-z)}{8\pi^2}\theta[L(s,s',q^2)],
\end{eqnarray}
where 
\begin{eqnarray}
L(s,s',q^2)=-m_b^2w-sw(w+z-1)-z[m_c^2-q^2w+s'(w+z-1)].
\end{eqnarray}

The QCD sum rules for the form factors are obtained by equating the hadronic and OPE sides of the correlator and applying the double Borel transformation with respect to the 
variables $p^2$ and $p^{'2}$ ($p^2 \rightarrow M_1^2$ and $p^{'2} \rightarrow M_2^{2}$ ). As a result we have
\begin{eqnarray}\label{borel}
f_{1[2]} (q^2) =& & -\frac{(m_b+m_u)}{f^*_B m^{*2}_B}\frac{(m_c+m_u)}{f^*_D m^{*2}_D}
e^{m^{*2}_B/M_1^2}e^{m^{*2}_D/M_2^2} \nonumber \\
 & &  \Big[ \int_{(m_b+m_u)^2}^{s_0} ds \int_{(m_c+m_u)^2}^{s'_0} ds' \rho_{1[2]}(s,s',q^2)
 e^{-s/M_1^2}e^{-s'/M_2^2} +\widehat{\textbf{B}}\Pi^{non-pert}_{1[2]} (q^2) \Big],  \nonumber \\
\end{eqnarray}
where $\widehat{\textbf{B}}$ represents the double Borel
transformation.
 As an example, we only show the explicit expression for the function $\widehat{\textbf{B}}\Pi^{non-pert}_{1}(q^2)$, which is given by
  \begin{eqnarray}
&&\widehat{\textbf{B}}\Pi^{non-pert}_{1}(q^2)\nonumber \\ &=&\frac{1}{2}\exp\Big[-\frac{m_b^2}{M^2}-\frac{m_c^2}{M'^2}\Big]
\Big[(-2m_u-m_b-m_c)\langle\bar{q}q\rangle_{\rho_N}+(p_0+p'_0)\langle q^{\dag} q\rangle_{\rho_N}\Big] \nonumber \\
& + &  \int_0^1 dz\int_0^{1-z}dw \frac{\langle g_s^2 G^2\rangle_{\rho_N} \exp\Big[\frac{(1-w)(m_c^2 w+m_b^2 z)}{M^2 z (w+z-1)}\Big]}{192 M^6 \pi^2 (w-1)z^4(w+z-1)^5}\Bigg\{\delta\Big(\frac{1}{M^{\prime^2}}+\frac{(1-w)w}{M^{^2}z(w+z-1)}\Big) \nonumber \\
 &\times&\Bigg[16m_b^4\pi^4(w-1)^2z^4(w+z-1)+16m_b^3\pi^4(w-1)z^3 \nonumber \\
 &\times&\Bigg(m_u(w-1)^2(w+2z-1)+m_c(w+z-1)\Big((w-1)w+(w-1)z+z^2\Big)\Bigg)\nonumber \\
 &+&w(w+z-1)\Bigg(16m_c^4\pi^4(w-1)^2w^3+3m_cm_uM^2(w-1)z^2(w+z-1)^2\nonumber \\
 &+&m_c^2M^2(w-1)wz\Big((1+16\pi^4)(w-1)w+2(-1+w+8\pi^4w)z+2z^2\Big)\nonumber \\
 &+&M^2z^2(w+z-1)\Big(-(2M^2+q^2)(w-1)w+(2M^2-q^2)(w-1)z\nonumber \\
 &+&(2M^2-q^2)z^2\Big)\Bigg)+m_b\Bigg(16m_c^3\pi^4(w-1)wz^2(w+z-1)\Big(w^2+w(z-1)+(z-1)z\Big)\nonumber \\
&+&8m_cM^2\pi^4z(w+z-1)\Big((w-1)^2w^3-6(w-1)^2wz^2-(w-1)(11w-5)z^3\nonumber \\
&-&10(w-1)z^4-5z^5\Big)+16m_c^2m_u\pi^4(w-1)^3w\Big(w^3+w^2(z-1)+wz^2+z^2(2z-1)\Big)\nonumber \\
&+&3m_uM^2(w-1)^2z^2(w+z-1)\Big(w^2+16\pi^4(1-2z)z+w(-1+z-16\pi^4z)\Big)\Bigg)\nonumber \\
&+&m_b^2(w-1)z(w+z-1)\Bigg(16m_cm_u\pi^4(w-1)z^2(w+z-1)+16m_c^2\pi^4(w-1)w(w^2+z^2)\nonumber \\
&+&M^2z\Big(w^3+16\pi^4(z-1)z^2+w^2(2z-1)+2wz(-1+z+8\pi^4z)\Big)\Bigg)\Bigg]\nonumber \\
&+&w\delta^{\prime}\Big(\frac{1}{M^{\prime^2}}+\frac{(1-w)w}{M^{^2}z(w+z-1)}\Big) \Bigg[-16m_c\pi^4(w-1)\Bigg(m_u(w-1)w^2(w+z-1)^2\nonumber \\
&+&m_b\Big(w^2+w(z-1)+(z-1)z\Big)\Big((w-1)w^2-(w-1)z^2-z^3\Big)\Bigg)\nonumber \\
&+&z\Bigg(16m_bm_u\pi^4(w-1)^3z(-1+z+2w)+M^2(w+z-1)\big(w^2+w(z-1)+(-1+z)z\Big)\nonumber \\
&\times&\Big(w^2+z-z^2-w(z+1)\Big)\Bigg)\Bigg]\Bigg\} \nonumber \\
&+&\frac{1}{12M^2 M^{\prime^2}}\exp\Bigg[-\frac{m_b^2}{M^2}-\frac{m_c^2}{M^{\prime^2}}\Bigg]\Bigg\{-\langle \bar{q}g_s\sigma
Gq\rangle_{\rho_N}\Big[m_c\Big(3M^2+M^{\prime^2}\Big)+m_b\Big(M^2+3M^{\prime^2}\Big)\Big] \nonumber \\
&+&\langle g_s\bar{q}\!\not\! {u}\sigma \cdot Gq\rangle_{\rho_N}\Big (M^2 p_0 +3M^{\prime^2} p_0+3M^2p^{\prime}_0+M^{\prime^2}p^{\prime}_0\Big)-4\Big(m_b M^2+m_cM^{\prime^2}\Big)\Big(\langle \bar{q}u\cdot D u\cdot Dq\rangle_{\rho_N} \nonumber \\
&+&im_q\langle \bar{q}\!\not\! {u} u\cdot Dq\rangle_{\rho_N}\Big)\Bigg\}
 \end{eqnarray}

\section{Numerical results}
In performing the numerical analysis of the sum rules for the form
factors $f_1(q^2)$ and $f_2(q^2)$, we need the values of some  input
parameters in nuclear medium entering into the sum rules. We present them in Table 1.
\begin{table}[ht!]
\centering
\rowcolors{1}{lightgray}{white}
\begin{tabular}{cc}
\hline \hline
   Input parameters  &  Values
           \\
\hline \hline
$m_{B}   $          &  $(5279.26\pm0.17)  $ $MeV$      \\
$m_{D}   $          &  $(1864.84 \pm 0.07)  $ $MeV$      \\
$p_0   $          &  $m_{B}  $      \\
$p'_0   $          &  $m_{D}$      \\
$ m_{u}   $          &  $2.3  $ $MeV$       \\
$ m_{d}   $          &  $4.8  $ $MeV$       \\
$ m_{b}   $          &  $4.18  $ $GeV$       \\
$ m_{c}   $          &  $1.275  $ $GeV$       \\
$ \rho_{N}     $          &  $(0.11)^3  $ $GeV^3$        \\
$ \langle q^{\dag}q\rangle_{\rho_N}    $          &  $\frac{3}{2}\rho_{N}$         \\
$ \langle\bar{q}q\rangle_{0}           $          &  $ (-0.241)^3    $ $GeV^3$          \\
$ m_{q}      $          &  $0.5(m_{u}+m_{d})$                 \\
$ \sigma_{N}            $          &  $0.045 ~  $GeV$ $                  \\
$  \langle\bar{q}q\rangle_{\rho_N}  $          &  $ \langle\bar{q}q\rangle_{0}+\frac{\sigma_{N}}{2m_
{q}}
\rho_{N}$                  \\
$  \langle q^{\dag}g_{s}\sigma
Gq\rangle_{\rho_N}  $          &  $ -0.33~GeV^2 \rho_{N}$                  \\
$  \langle q^{\dag}iD_{0}q\rangle_{\rho_N}  $          &  $0.18 ~GeV \rho_{N}$                  \\
$  \langle\bar{q}iD_{0}q\rangle_{\rho_N}  $          &  $\frac{3}{2} m_q \rho_{N}\simeq0 $                  \\
$  m_{0}^{2}  $          &  $ 0.8~GeV^2$                  \\
$   \langle\bar{q}g_{s}\sigma Gq\rangle_{0} $          &  $m_{0}^{2}\langle\bar{q}q\rangle_{0} $                  \\
$  \langle\bar{q}g_{s}\sigma
Gq\rangle_{\rho_N}  $          &  $\langle\bar{q}g_{s}\sigma Gq\rangle_{0}+3~GeV^2\rho_{N} $                  \\
$ \langle  \bar{q}iD_{0}iD_{0}q\rangle_{\rho_N} $          &  $ 0.3~GeV^2\rho_{N}-\frac{1}{8}\langle\bar{q}g_{s}
\sigma
Gq\rangle_{\rho_N}$                  \\
$  \langle
q^{\dag}iD_{0}iD_{0}q\rangle_{\rho_N}  $          &  $0.031~GeV^2\rho_{N}-\frac{1}{12}\langle
q^{\dag}g_{s}
\sigma Gq\rangle_{\rho_N} $                  \\
$\langle \frac{\alpha_s}{\pi} G^{2}\rangle_{0}$ & $(0.33\pm0.04)^4~GeV^4$\\
$\langle \frac{\alpha_s}{\pi} G^{2}\rangle_{\rho_N}$ & $\langle \frac{\alpha_s}{\pi} G^{2}\rangle_{0}-0.65~GeV \rho_N$\\
 \hline \hline
\end{tabular}
\caption{Numerical values for input parameters \cite{cohen95,Nielsen,XJ1,cohen45,PDG}. The value presented for $\rho_N$ corresponds to the nuclear matter saturation density which is used in numerical calculations. }
\end{table}
Besides these input parameters, the sum rules for the form factors
contain four auxiliary parameters, viz. the Borel mass
parameters $M^2$ and $M'^2$ as well as continuum thresholds $s_0$ and
$s'_0$. The  physical quantities like form factors should be
roughly independent of these parameters according to the general philosophy of the method used.
In the following, we shall find their working regions such that
the values of form factor weakly depend on these parameters.

 The
continuum thresholds are not entirely capricious but they depend on
the energy of the first excited states in the initial and final channels with the same quantum
numbers as the interpolating currents. From numerical analysis, the working intervals are
obtained as $s_0=(32.0\pm1.5)~GeV^2$ and $s'_0=(5.0\pm0.5)~GeV^2$ for the continuum thresholds.  The Borel mass parameters are restricted by requirements that,
 not only the contributions of the higher states and continuum are  sufficiently suppressed but also the contributions of the higher dimensional operators are small.  
These conditions lead to   the intervals
$8~GeV^2\leq M^2\leq 12~GeV^2$ and $4~GeV^2\leq M^2\leq 6~GeV^2$. In order to see how our results depend on the Borel
parameters, we present the dependence of the form factors
$f_1(0)$ and $f_2(0)$, at fixed values of the continuum thresholds, on these parameters in
Figs. 1 and 2.   In these figures, the solid lines  stand for the nuclear matter results and dashed lines for those of vacuum. 
From these figures we see that not only the form factors demonstrate good stabilities with respect to the variations of Borel parameters in their working regions, but the  results obtained in the nuclear
medium differ considerably with those of the vacuum. 
%The shifts are large in the case of $f_1$ compared to the form factor $f_2$.
\begin{figure}[h]
\centering
\begin{tabular}{cc}
\includegraphics[totalheight=6cm,width=7cm]{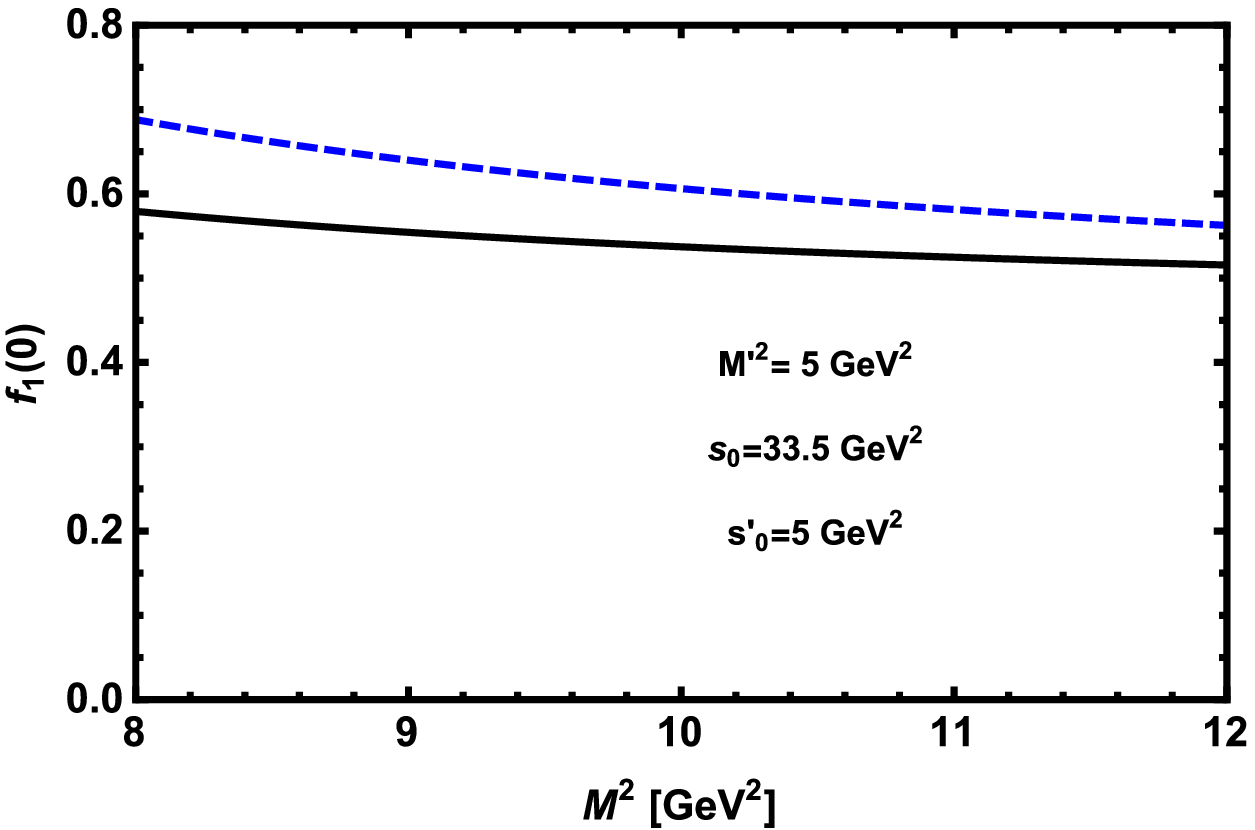}
\includegraphics[totalheight=6cm,width=7cm]{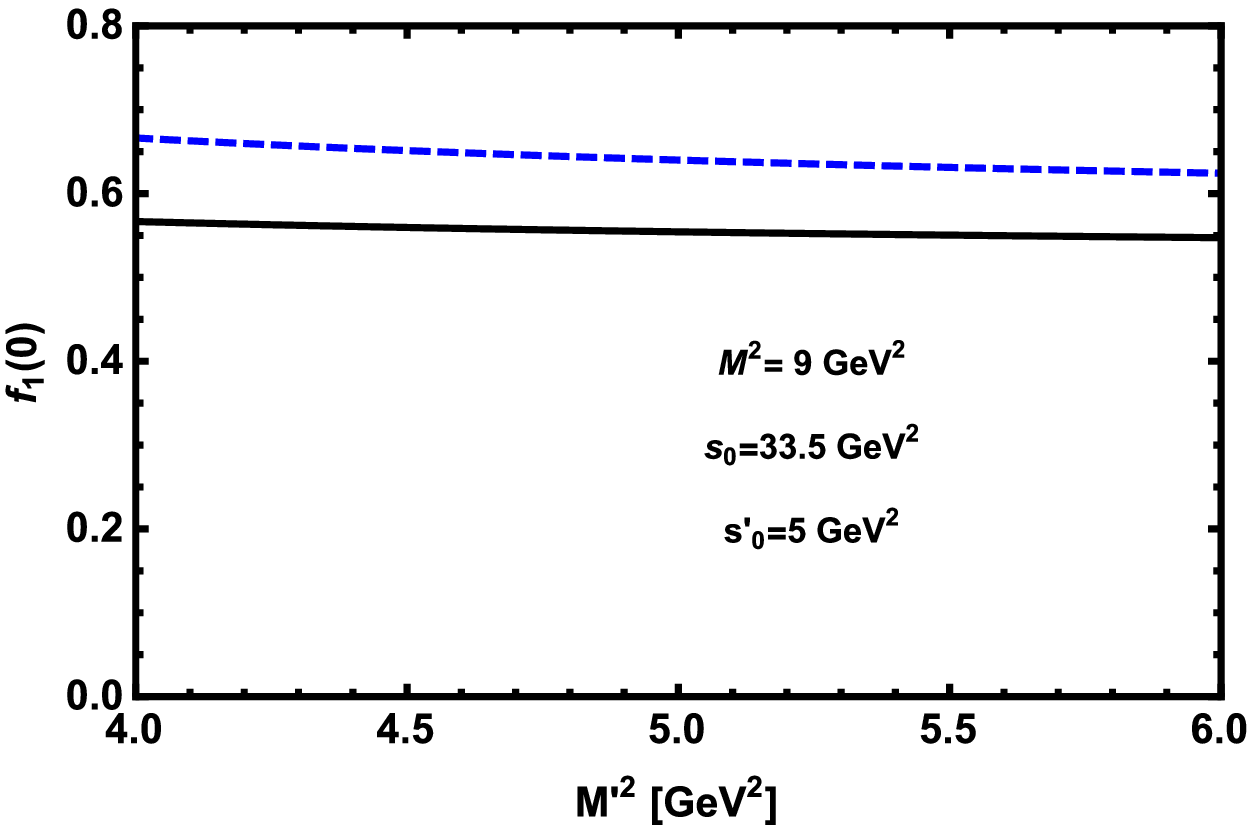}
\end{tabular}
\caption{Left panel: the form factor $f_1(0)$ versus $M^2$ in nuclear matter  (solid line) and  vacuum  (dashed line). 
 Right panel: the form factor $f_1(0)$ versus $M'^2$ in nuclear matter (solid line)  and vacuum (dashed line). }
\end{figure}
\begin{figure}[h]
\centering
\begin{tabular}{cc}
\includegraphics[totalheight=6cm,width=7cm]{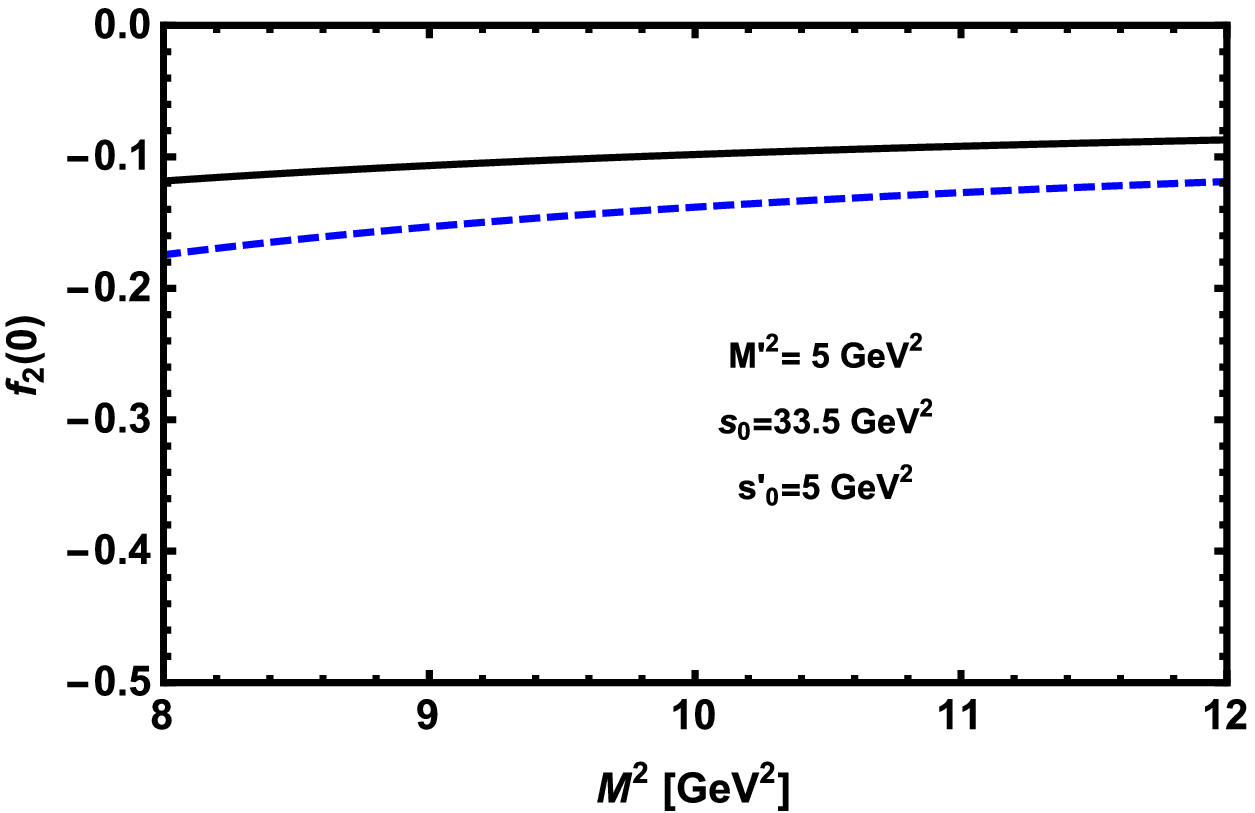}
\includegraphics[totalheight=6cm,width=7cm]{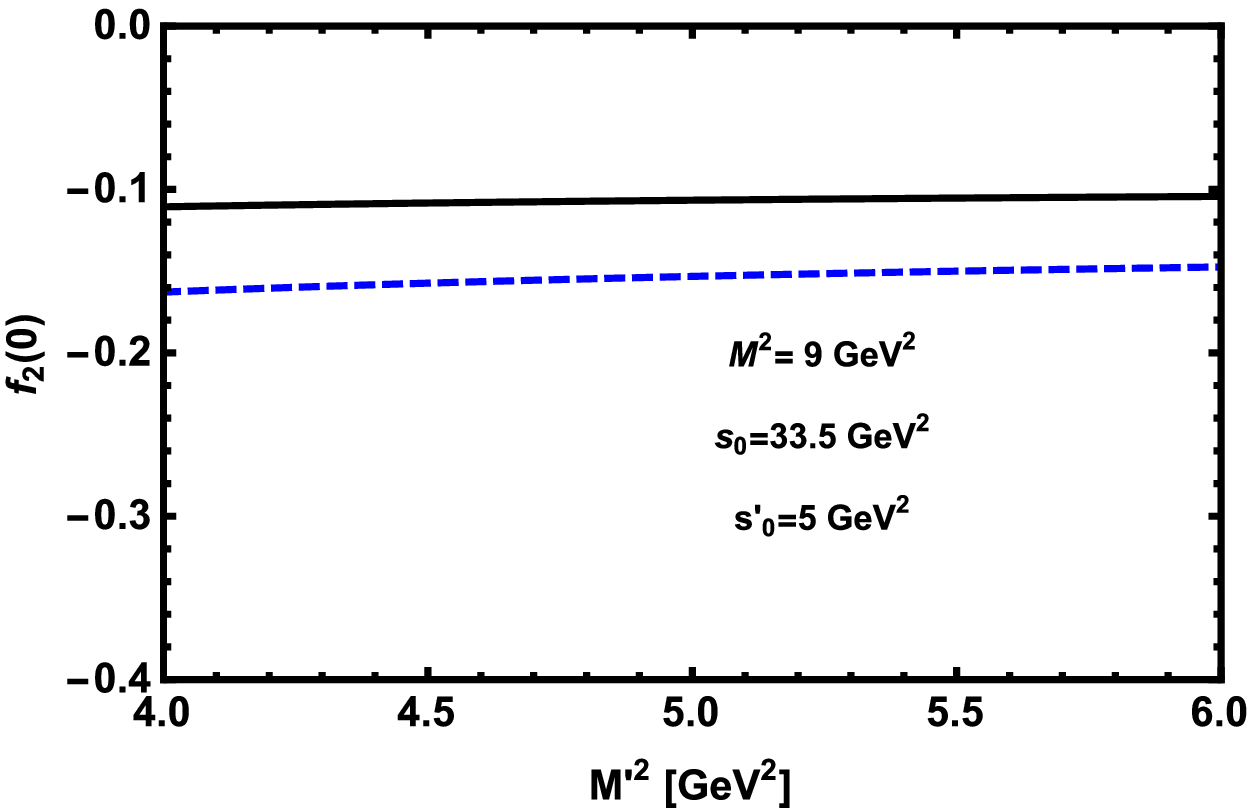}
\end{tabular}
\caption{The same as figure $1$ but for  $f_2(0)$ form factor.}
\end{figure}
\begin{figure}[h]
\centering
\begin{tabular}{cc}
\includegraphics[totalheight=6cm,width=7cm]{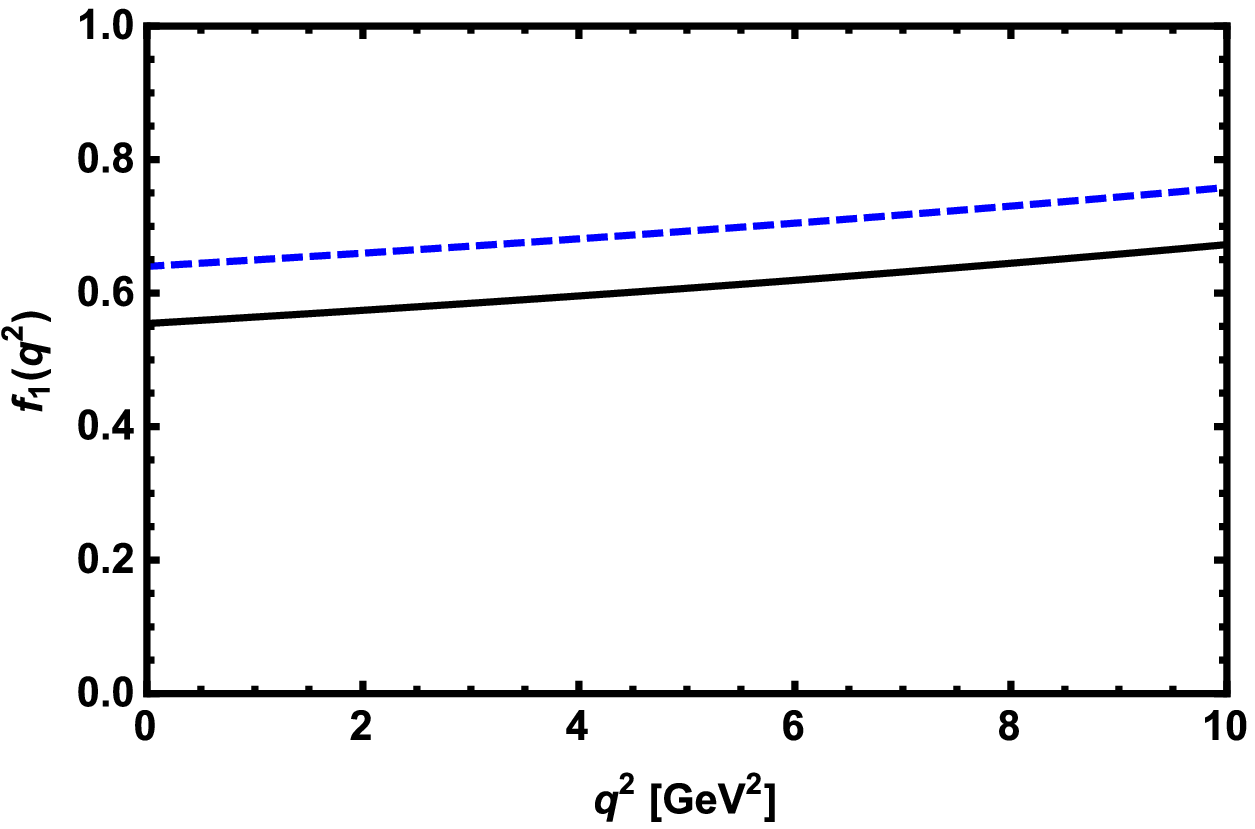}
\includegraphics[totalheight=6cm,width=7cm]{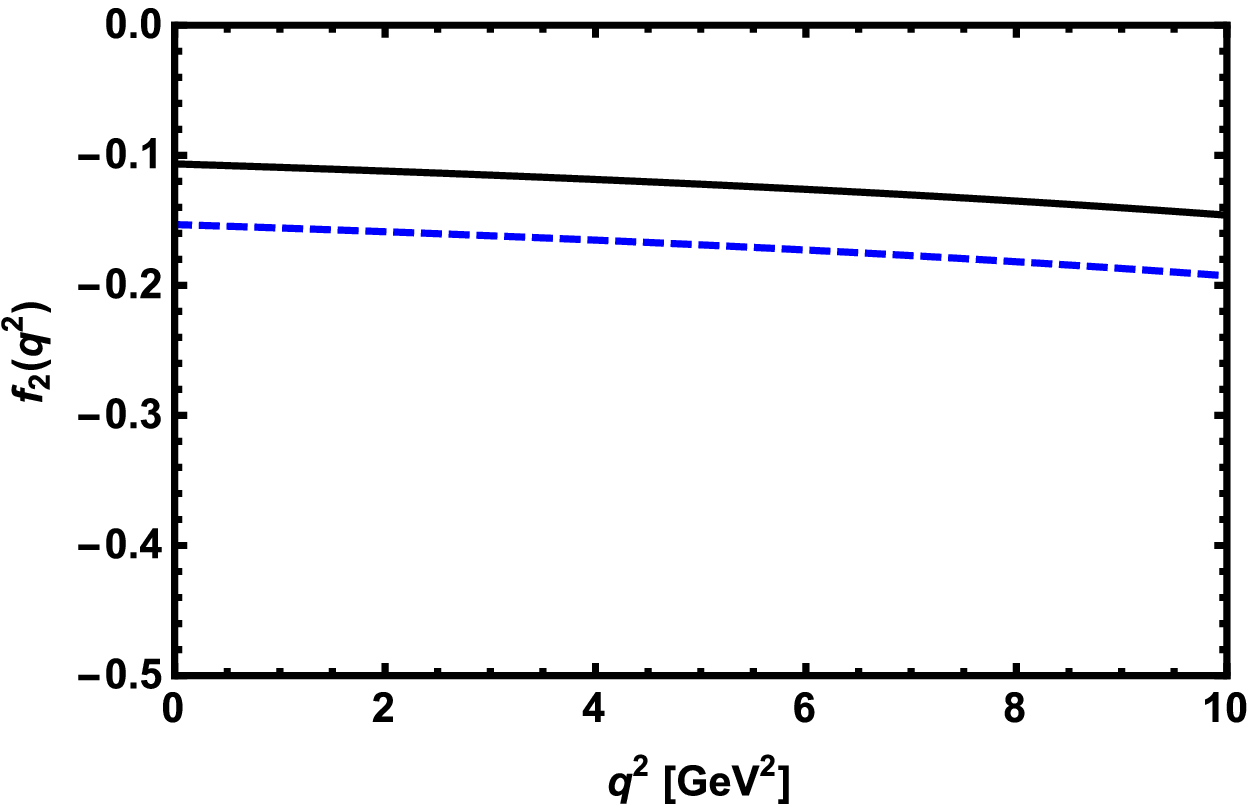}
\end{tabular}
\caption{The  dependence of the form factors $f_1(q^2)$ (left panel) and $f_2(q^2)$ (right panel) on $q^2$ in nuclear matter (solid line) and  vacuum (dashed line).   }
\end{figure}

Having determined the working regions for the  continuum
thresholds and Borel mass parameters, we proceed to find the
behaviors of the form factors in terms of $q^2$ . Our analysis
shows that the form factors are well fitted to the following
function:

\begin{equation}
\label{ }
f_i(q^2) = \frac{f_i(0)}{1+\alpha \hat{q} +\beta \hat{q}^2 +\gamma \hat{q}^3 +\lambda \hat{q}^4},
\end{equation}
where $\hat{q}=q^2/m^2_{B_q}$  and the numerical values for the parameters $f_i(0)$, $\alpha$, $\beta$, $\gamma$ and $\lambda$  are presented in tables 2 and 3 for nuclear matter and vacuum, respectively.
From these tables we see that although the central values of the form factors at $q^2=0$ seem to be  considerably different in medium and vacuum, considering the errors roughly  kills  these differences. 
The errors in the results belong to the uncertainties in
determination of the working regions for the auxiliary parameters
as well as the errors in the other input parameters.

\begin{table}[ht!]
\centering
\rowcolors{1}{lightgray}{white}\begin{tabular}{c|c|c|c|c|c}\hline\hline
& $f(0)$ & $\alpha$ & $\beta$ & $\gamma$ & $\lambda$ \\\hline\hline
$f_1$ &  $0.55 \pm 0.06$ & $-0.48$ & $-0.036 $  & $0.008$  & $0.007$ \\
$f_2 $& $-0.11 \pm 0.01$ & $-0.67$ & $-0.247$ & $0.006$  & $0.079$ \\
\end{tabular}
\caption{The numerical values for the parameters $f_i(0)$, $\alpha$, $\beta$, $\gamma$ and $\lambda$ in nuclear matter.}
\end{table}
\begin{table}[ht!]
\centering
\rowcolors{1}{lightgray}{white}\begin{tabular}{c|c|c|c|c|c}\hline\hline
& $f(0)$ & $\alpha$ & $\beta$ & $\gamma$ & $\lambda$ \\\hline\hline
$f_1$ &  $0.64 \pm 0.07$ & $-0.41$  & $-0.057$ & $0.002$ & $0.006$ \\
$f_2 $&  $-0.15 \pm 0.02$ & $-0.47$ & $-0.266$ &$-0.085$ & $0.023$ \\
\end{tabular}
\caption{The numerical values for the parameters $f_i(0)$, $\alpha$, $\beta$, $\gamma$ and $\lambda$ in vacuum.}
\end{table}
We  present the dependence of $f_1(q^2)$ and $f_2(q^2)$ on $q^2$ at average values of the Borel mass parameters and continuum thresholds in figure 3. From this figure we also see that, as far as the central values
 are concerned, the nuclear medium affect
the $q^2$ dependencies of the form factors considerably.

The next step is to calculate the branching ratio of
the process under consideration both in nuclear medium and vacuum for both $\ell=e,\mu$ and $\tau$, which are depicted in tables 4 and 5. For comparison, 
we also depict the existing experimental data.
From these tables we obtain that the nuclear medium suppresses the values of branching ratios with amount of roughly 50\% for all lepton channels when the central values are considered. It is also seen that the 
errors can not kill the differences between the medium and vacuum predictions in the case of branching fractions.
The results of vacuum sum rules are consistent with  the experimental data \cite{PDG} for all lepton channels within the errors.
%\begin{eqnarray}\label{29au}
%\frac{d\Gamma}{dq^2}&=&\frac{1}{192\pi^{3}m_{B_{q}}^{3}}
%G_{F}^2|V_{cb}|^2\lambda^{1/2}(m_{B_{q}}^{2},m_{D_{q}}^{2},q^{2})\left(\frac{q^{2}-m_{\ell'}^{2}}{q^{2}}\right)^{2}
%\nonumber \\
%&\times&\left\{-\frac{1}{2}(2q^{2}+m_{\ell}^{2})\left[|f_{1}(q^{2})|^{2}(2m_{B_{q}}^{2}+2m_{D_{q}}^{2}-q^{2})
%\right.\right. \nonumber \\
%&+&\left.\left.2(m_{B_{q}}^{2}-m_{D_{q}}^{2})Re[f_{1}(q^{2})f_{2}^{*}(q^{2})]+|f_{2}(q^{2})|^{2}q^{2}\right]\right.\nonumber
%\\
%&+&\left.\frac{(q^{2}+m_{\ell'}^{2})}{q^{2}}\left[|f_{1}(q^{2})|^{2}(m_{B_{q}}^{2}-m_{D_{q}}^{2})^{2}
%\right.\right. \nonumber \\
%&+&\left.\left.2(m_{B_{q}}^{2}-m_{D_{q}}^{2})q^{2}Re[f_{1}(q^{2})f_{2}^{*}(q^{2})]+|f_{2}(q^{2})|^{2}q^{4}\right]
%\right\}.
%\end{eqnarray}

%After performing integration over $q^2$  in the interval
%$m^2_{\ell'}\leq q^2 \leq (m_B-m_D)^2$, we obtain the branching ratios for leptons $\ell=e, \mu$ and $\tau$ using three different fit
%functions presented in Table 4 for nuclear matter and in Table 5 for vacuum.

\begin{table}[ht!]
\centering
\rowcolors{1}{lightgray}{white}\begin{tabular}{l | l | l}\hline\hline
$Br $& $B^{+} \rightarrow \overline{D}^{0} \ell^{+}\nu_{\ell}$ & $B^{+} \rightarrow \overline{D}^{0} \tau^{+}\nu_{\tau}$  \\\hline\hline
%$\Gamma \textrm{(GeV)}$ &  $1.015 \times 10^{-14}$  & $1.014 \times 10^{-14}$   & $0.424\times 10^{-14}$    \\
Nuclear matter & ($1.04 \pm 0.29$)$ \times 10^{-2}$  &($0.48 \pm 0.13$)$\times 10^{-2}$ 
\end{tabular}
\caption{The branching ratios  in nuclear matter.}
\end{table}
\begin{table}[ht!]
\centering
\rowcolors{1}{lightgray}{white}\begin{tabular}{l | l | l}\hline\hline
$Br $& $B^{+} \rightarrow \overline{D}^{0} \ell^{+}\nu_{\ell}$ & $B^{+} \rightarrow \overline{D}^{0} \tau^{+}\nu_{\tau}$  \\\hline\hline
%$\Gamma \textrm{(GeV)}$ &  $1.015 \times 10^{-14}$  & $1.014 \times 10^{-14}$   & $0.424\times 10^{-14}$    \\
%Nuclear matter & ($1.91 \pm 0.53$)$ \times 10^{-2}$  &($0.8 \pm 0.022$) $\times 10^{-2}$   \\
Vacuum &($2.03 \pm 0.57$) $ \times 10^{-2}$ &($0.92 \pm 0.26$) $\times 10^{-2}$  \\
%pQCD \cite{Ying}& $2.19^{+0.99}_{-0.76}$ $ \times 10^{-2}$&$0.95^{+0.37}_{-0.31}$ $ \times 10^{-2}$ \\
PDG \cite{PDG}& ($2.27 \pm 0.11$)$ \times 10^{-2}$  & ($0.77 \pm 0.25$)$ \times 10^{-2}$ \\
%BaBar\cite{BaBar2008,BaBar2009} & ($2.34 \pm 0.03 \pm 0.13$)$ \times 10^{-2}$  & ($0.67 \pm 0.37 \pm 0.11 \pm 0.07$)$ \times 10^{-2}$ \\
\end{tabular}
\caption{The branching ratios  in vacuum together with the experimental data.}
\end{table}

%In Table 4 and 5 instead of the branching ratios of electron and muon used only the branching ratio of $\ell$. 
%IThere are two reason for this: the first one is the numerical value of the branching ratio of $e$ and $\mu$ are
%I very close, the second one is that in the literature the theoretical and the experimental results of vacuum are 
%Irepresented in the same way, as seen in Table 5. Our vacuum results of the branching ratio of $B^{-} \rightarrow D^{0} \ell\bar{\nu}_{\ell}$ 
%Iand $B^{-} \rightarrow D^{0} \tau\bar{ \nu}_{\tau}$ decays are consistent with experimental BaBar result and the theoreticalresults of the perturbative QCD \cite{Ying} and with PDG one.

%The errors in the results belong to the uncertainties in
%determination of the working regions for the auxiliary parameters
%as well as errors in the other input parameters.

At the end of this section, we would like to calculate the  ratio of branching fractions in $\tau$ to $\ell=\mu, e$ channels, i.e.
\begin{eqnarray}
\ensuremath{{\cal R}(D)} =\frac{{\cal B}(B^+\rightarrow \overline{D}^0
\tau^{+}\nu_{\tau})}{{\cal B}(B^+\rightarrow \overline{D}^0
\ell^{+}\nu_{\ell})},
\end{eqnarray}
for both the nuclear medium and vacuum. We obtain the value $\ensuremath{{\cal R}(D)}=0.461\pm 0.009$  for the nuclear medium which is roughly the same with the
  vacuum value $\ensuremath{{\cal R}(D)}=0.453\pm 0.009$ within the errors.

We conclude that although the nuclear medium effects cause considerable shifts in the central values of the form factors, considering the errors roughly kills these differences.
In the case of  branching ratios we see considerable differences between the medium and vacuum predictions for all lepton channels, which can not be killed by the errors of the form factors. This can be attributed to the 
shifts in the masses of the participating mesons due to the nuclear medium. The  ratio of branching fractions in $\tau$ to $\ell=\mu, e$ channel remains roughly
unchanged both in medium and vacuum. This ratio and other quantities in nuclear medium considered in the present work can be checked in future in-medium experiments.

\section{Acknowledgement}
This work has been supported in part by the Scientific and Technological Research Council of Turkey (TUBITAK) under the research project 114F018.

                                 %%%%%%%%%%%%%%%%%%%%%%%%%%%%%%%%%%%%%%%
       %%%%%%%%%%%%%%%%%%%%%%%%%%%%%%%%%%%%%%%             %%%%%%%%%%%%%%%%%%%%%%%%%%%%%%%%%%%%%%%%%%
                                 %%%%%%%%%%%%%%%%%%%%%%%%%%%%%%%%%%%%%%%

\begin{thebibliography}{99}

 \bibitem{BABAR} J. P. Lees et al. (BABAR Collaboration), Phys. Rev. Lett. 109, 101802 (2012).
\bibitem{colangelo} P. Biancofiore, P. Colangelo, F. De Fazio, Phys. Rev. D 87, 074010 (2013).
\bibitem{fajfer}  S. Fajfer, J. F. Kamenik and I. Nisandzic, Phys. Rev. D 85, 094025 (2012).
\bibitem{fajfer1} S. Fajfer, J. F. Kamenik, I. Nisandzic and J. Zupan, Phys. Rev. Lett. 109, 161801 (2012).
\bibitem{Crivellin} A. Crivellin, C. Greub and A. Kokulu, Phys. Rev. D 86, 054014 (2012).
\bibitem{Datta} A. Datta, M. Duraisamy and D. Ghosh, Phys. Rev. D 86, 034027 (2012).
\bibitem{Becirevic} D. Becirevic, N. Kosnik and A. Tayduganov, Phys. Lett. B 716, 208 (2012).
\bibitem{damir1} D. Becirevic, N. Kosnik and A. Tayduganov, PoS ConfinementX, 244 (2012). 
\bibitem{Celis} A. Celis, M. Jung, X. -Q. Li and A. Pich, JHEP 1301, 054 (2013).
\bibitem{Choudhury} D. Choudhury, D. K. Ghosh and A. Kundu, Phys. Rev. D 86, 114037 (2012).
\bibitem{Tanaka} M. Tanaka and R. Watanabe,  Phys. Rev. D 87, 034028 (2013).
\bibitem{Bhol} A. Bhol, EPL, 106, 31001 (2014).



%\bibitem{BABAR1} J. P. Lees et al., BABAR Collaboration, Phys. Rev. D 88, 072012 (2013).
%\bibitem{LHCb} A. Aaij et al., LHCb Collaboration, J. High Energy Phys. 08, 131  (2013).
%\bibitem{LHCb1} A. Aaij et al., LHCb Collaboration, Phys. Rev. Lett. 111, 191801 (2013).
%\bibitem{Azizi} K. Azizi, Nucl. Phys. B 801, 70 (2008).
%\bibitem{Fu} H.-B. Fu, X.-G. Wu, H.-Y. Han, Y. Ma, Nucl. Phys. B 884,
%172  (2014).
%\bibitem{Azizi1} K. Azizi, H. Sundu, S. Sahin,  Phys. Rev. D 88, 036004  (2013).
%\bibitem{Segovia} J. Segovia, E. Hern$\acute{a}$ndez, F. Fern$\acute{a}$ndez, D. R. Entem, Phys. Rev. D 87, 114009  (2013).
%\bibitem{Segovia1} J. Segovia, C. Albertus, E. Hern$\acute{a}$ndez, F. Fern$\acute{a}$ndez, D. R. Entem,  Phys. Rev. D 86, 014010  (2012).
% \bibitem{Fu1} H.-F. Fu, G.-L. Wang, Z.-H. Wang, X.-J. Chen, Chin.
 %Phys. Lett. Vol. 28, No. 12, 121301 (2011).
 %\bibitem{Ivanov} M. A. Ivanov, J. G. Korner, S. G. Kovalenko, P.
% Santorelli, G. G. Saidullaeva,  Phys. Rev. D 85, 034004  (2012).
% \bibitem{Ebert} D. Ebert, R. N. Faustov, V. O. Galkin, Phys. Rev. D 85, 054006  (2012).
 %\bibitem{Katirci} N. Kat{\i}rc{\i}, K. Azizi, JHEP 1107, 043 (2011) .
% \bibitem{Damir} D. Be\v{c}irevi\'{c}, N. Ko\v{s}nik, A.
%Tayduganov, Phys. Lett. B 716, 208  (2012).

\bibitem{Hayashigaki} A. Hayashigaki, Phys. Lett. B 487, 96 (2000).
\bibitem{Hilger} T. Hilger, R. Thomas, B. K\"ampfer, Phys. Rev. C 79, 025202 (2009).
\bibitem{Hilger2} T. Hilger, B. K\"ampfer,  Nucl. Phys. Proc. Suppl. 207-208, 277 (2010).
\bibitem{Wang2011} Z.-G. Wang, T. Huang, Phys. Rev. C 84, 048201 (2011).
\bibitem{Azizi2} K. Azizi, N. Er, H. Sundu, Eur. Phys. J. C 74, 3021 (2014).

\bibitem{bir}  E. Fioravanti, arXiv:1206.2214.
\bibitem{iki}  B. Friman et al, ``The CBM physics book: Compressed Baryonic Matter in Laboratory Experiments'', Springer Heidelberg.
\bibitem{uc} http://www.gsi.de/fair/experiments/CBM/index e.html.
\bibitem{dort} http://www-panda.gsi.de/auto/phy/ home.htm.

\bibitem{cohen95}T. D. Cohen, R. J. Furnstahl, D. K. Griegel, X. Jin, Prog. Part. Nucl.
Phys. 35, 221 (1995).
\bibitem{reinders85}L. J. Reinders, H. Rubinstein, S. Yazaki, Phys. Rep. 127, 1 (1985).
\bibitem{XJ1}X. Jin, T. D. Cohen, R. J. Furnstahl, and D. K. Griegel, Phys. Rev. C 47, 2882 (1993).
 \bibitem{Nielsen}X. Jin, M. Nielsen, T. D. Cohen, R. J. Furnstahl, D. K. Griegel, Phys. Rev. C 49, 464 (1994).
 \bibitem{cohen45}T. D. Cohen, R. J. Furnstahl and D. K. Griegel, Phys. Rev. C 45, 1881 (1992).
\bibitem{PDG} J. Beringer  et al., Particle Data Group,  Phys. Rev. D 86, 010001 (2012).
% \bibitem{Ying}Y.-Y. Fan, W.-F. Wang, S. Cheng, Z.-J. Xiao, Chin. Sci. Bull 59, 125  (2014).
 % \bibitem{BaBar2008} B. Aubert  et al.( BABAR Collaboration), Phys. Rev. Lett. 100, 021801  (2008).
% \bibitem{BaBar2009} B. Aubert  et al. (BABAR Collaboration), Phys. Rev. D 79, 012002 (2009).
 %\bibitem{Ball91}P. Ball, V. M. Braun, and H. G. Dosch, Phys. Rev. D44, 3567  (1991).
% \bibitem{Ball93}P. Ball, Phys. Rev. D 48, 3190  (1993).

% \bibitem{CMS} S. Chatrchyan et al. [CMS Collaboration], Phys. Lett. \textbf{B 716}, (2012)  30 [arXiv:1207.7235 [hep-ex]].

%\bibitem{ATLAS} G. Aad et al. [ATLAS Collaboration], Phys. Lett. \textbf{B 716}, (2012) 1  [arXiv:1207.7214 [hep-ex]]. G. Aad et al. [ATLAS Collaboration], Phys. Rev.\textbf{ D 86}, (2012) 032003  [arXiv:1207.0319 [hep-ex]].

%\bibitem{Dutta}    R. Dutta, A. Bhol, A. K. Giri,  Phys. Rev. \textbf{D 88}, (2013) 114023 .



%(**************************************************************************)


%%
\end{thebibliography}
\end{document}